\documentclass{mpe_report}

\usepackage{psfig,graphicx,epsfig}
\usepackage{color}
\usepackage{amsmath,amssymb,epic,eepic,array}

\begin{document}

\pagenumbering{arabic}
\setcounter{page}{145}

 \renewcommand{\FirstPageOfPaper }{145}\renewcommand{\LastPageOfPaper }{148}%\documentclass{mpe_report}
%\usepackage{psfig}
%\def\R{~ROSAT}
%\def\RAS{\R all sky survey}
%\begin{document}

\title{Magnetised Electron-Positron Plasmas}
\author{C.R. Stark\inst{1}, D.A. Diver\inst{1} \and A.A. da Costa\inst{2}}
\institute{Dept. of Physics and Astronomy, University of Glasgow,
Glasgow G12 8QQ, Scotland, UK \and Sec\c{c}\~{a}o de
Telecomunica\c{c}\~{o}es, DEEC/SPR, Instituto Superior T\'{e}cnico,
1049-001 Lisboa, Portugal}

\maketitle

\begin{abstract}
 Electrostatic oscillations in cold electron-positron plasmas can be
 coupled to a propagating electromagnetic mode if the background
 magnetic field is inhomogeneous.  Previous work considered this
 coupling in the quasi-linear regime, successfully simulating the
 electromagnetic mode.  Here we present a stability analysis of the
 non-linear problem, perturbed from dynamical equilibrium, in order
 to gain some insight into the modes present in the system.
\end{abstract}

\section{Introduction}
Pulsar magnetospheres are composed of magnetised electron-positron
plasmas. Conventional modelling uses the fact that the dominant
Lorentz force produces a host of relativistic charged particles,
each of which radiates strongly and stochastically, producing
$\gamma$-ray photons. Such single-particle models have been explored
as possible radiation sources (da Costa \& Kahn \cite{costa97}), but
have not been able to recover the highest energy radiation observed.
\par
However this strict single particle approach may be challenged as
there are physical processes which develop in the rest frame of the
plasma, and are strongly amplified when translated in the laboratory
frame of the pulsar.  In cold non-relativistic plasma theory
non-linear electrostatic oscillations in electron-positron plasmas
develop a density instability in which the density of both species
grows sharply at the edges of the oscillation site (da Costa {\it et
al}. \cite{costa01}). Folding thermodynamics into the system
provides a possible mechanism for avoiding the onset of the
instability, since pressure effects would oppose the density build
up. However, a more advantageous mechanism would be coupling the
oscillation to an electromagnetic mode via an inhomogeneous
background magnetic field. This would allow energy to be radiated
away from the oscillation, quenching the density instability and
giving a source of radiation in the pulsar magnetosphere.  Previous
work has studied this mechanism and successfully simulated the
coupling between the oscillation and the electromagnetic mode in the
quasi-linear regime (Diver {\it et al}. \cite{diver}).  Therefore
the study of these processes is of the greatest importance for the
dynamics of the magnetosphere, and the radiation mechanisms of
pulsars.

This paper addresses the basic mathematical formulation of the
electron-positron cold magnetoplasma and the mode coupling mechanism
in Section 2.  In Section 3 linear analysis of the equations
perturbed from dynamical equilibrium is performed to gain an
understanding of the modes present in the plasma and their
stability.

\section{Model equations}
In cylindrical polar coordinates $\left(r,\theta,z\right)$ with an
axial magnetic field $\vec{B}=\vec{\hat z}B_{z}$ and an electric
field in the $r,\theta$ plane $E_{r,\theta}$ the full
non-linear model equations for a cold electron-positron plasma are
\begin{eqnarray}
r\dot{n}_+ +(rn_+u_r)'&=&0\\
 r\dot{n}_- +(rn_-v_r)'&=&0\label{nl1}\\
\dot{u}_r+u_ru_r'-u_{\theta}^2/r&=&(e/m)(E_r+u_{\theta}B_z)\\
\dot{u}_{\theta}+u_ru_r'+u_ru_{\theta}/r&=&(e/m)(E_{\theta}-u_rB_z)\\
\dot{v}_r+v_rv_r'-v_{\theta}^2/r&=&-(e/m)(E_r+v_{\theta}B_z)\\
\dot{v}_{\theta}+v_rv_r'+v_rv_{\theta}/r&=&-(e/m)(E_{\theta}-v_rB_z)\\
(rE_r)'&=&(e/\epsilon_0)r(n_+-n_-)\label{dive}\\
0&=&-\dot{E}_r/c^2-\mu_0e(n_+u_r-n_-v_r)\label{curlbr}\\
(rE_{\theta})'&=&-r\dot{B_z}\label{curle}\\
B_z'&=&-\dot{E}_{\theta}/c^2-\mu_0e(n_+u_{\theta}-n_-v_{\theta})\label{curlbt}
\end{eqnarray}
where $n_+$, $n_-$ are the positron and electron number densities;
$u$, $v$ are the positron and electron velocities; $\dot{}$ denotes
$\partial/\partial t$; $'$ denotes $\partial /\partial r$;
(\ref{dive}) is Poisson's equation; (\ref{curle}) is the single $z$
component of the induction equation; and (\ref{curlbt}),
(\ref{curlbr}) are the $\theta$ and $r$ components of the $\nabla
\times \vec{B}$ equation.  The equations can be recast into a form
that highlights the symmetry of the electron-positron plasma, this
can be done via the following relations
\begin{eqnarray}
\Sigma&=&\frac{1}{2}\left(n_{+}+n_{-}\right)\xi/n_{0} \\
\Delta&=&\frac{1}{2}\left(n_{+}-n_{-}\right)\xi/n_{0} \\
\sigma&=&\frac{1}{2}\left(u_{r}+v_{r}\right)/\omega_{0}L \\
\delta&=&\frac{1}{2}\left(u_{r}-v_{r}\right)/\omega_{0}L \\
\chi&=&\frac{1}{2}\left(u_{\theta}+v_{\theta}\right)/\omega_{0}L \\
\zeta&=&\frac{1}{2}\left(u_{\theta}-v_{\theta}\right)/\omega_{0}L \\
\rho&=&\frac{eE_{r}}{m\omega_{0}^{2}L}\\
\theta&=&\frac{eE_{\theta}}{m\omega_{0}^{2}L}\\
\beta_{0}+\beta&=&\frac{eB_{z}}{m\omega_{0}} \\
r&=&\xi L \\
t&=&\tau/\omega_{0}
\end{eqnarray}
where $L,1/\omega_{0}$ are characteristic length, time scales and
$\beta_{0}$, $\beta$ are the equilibrium and perturbed magnetic
field respectively. The system of governing equations then becomes
\begin{eqnarray}
\dot{\Sigma}&=&-\left(\Sigma \sigma + \Delta \delta \right)' \\
\dot{\Delta}&=&-\left(\Delta \sigma + \Sigma \delta \right)' \\
\dot{\sigma}&=&-\frac{1}{2}\left(\sigma^{2}+\delta^{2}
\right)'+\left(\chi^{2}+\zeta^{2}\right)/\xi+\zeta\left(\beta_{0}+\beta\right) \\
\dot{\delta}&=&-\left( \sigma \delta
\right)'+2\chi\zeta/\xi+\rho+\chi\left(\beta_{0}+\beta\right) \\
\dot{\chi}&=&-\chi'\sigma-\zeta'\delta-\left(\chi\sigma+\zeta\delta\right)/\xi-\delta
\left(\beta_{0}+\beta\right)\\
\dot{\zeta}&=&-\sigma\zeta'-\delta\chi'-\left(\sigma\zeta+\delta\chi\right)/\xi+\theta-\sigma
\left(\beta_{0}+\beta\right)\\
\dot{\rho}&=&-\frac{2}{\xi}\left(\Delta\sigma+\Sigma\delta\right)\\
\dot{\theta}&=&-p\left(\beta_{0}+\beta\right)'-\frac{2}{\xi}\left(\Sigma\zeta+\Delta\chi\right)\\
\dot{\beta}&=&-\theta'-\theta/\xi
\end{eqnarray}
where $p=c^{2}/(\omega_{0}^{2}L^{2})$.
\subsection{The coupling mechanism}
Consider a homogeneous background magnetic field.  When the density
of the plasma is perturbed, resulting in a charge imbalance, a
radial electric field is created that accelerates the electrons and
positrons in radially opposite directions.  The plasma collectively
is responding to the presence of the electric field and is trying to
restore equilibrium.  The magnetic field causes the particle
trajectories of both species to participate in partial Larmor orbits
with the same azimuthal velocity.  The particles overshoot their
initial positions, due to their acquired kinetic energy, and produce
a new charge imbalance that they again try to correct.  This is one
half period of an electrostatic oscillation.

Introducing an inhomogeneous magnetic field induces a
$\vec{B}\times\nabla B$ drift which causes a net current density in
the azimuthal motion of the plasma during the electrostatic
oscillation.  The resulting current density induces axial magnetic
field fluctuations which propagate away from the electrostatic
oscillation site.  This mechanism causes the electrostatic
oscillations to couple to a propagating electromagnetic mode in the
plasma.

This effect has been fully investigated in the quasi-linear regime,
in which the variables were perturbed from a uniform equilibrium.
However having an inhomogeneous magnetic field permeating the plasma
requires the equilibrium to be non-uniform for the system to be
self-consistent.

\subsection{Dynamical equilibrium}
In the equilibrium situation $\partial/\partial \tau=0$ and
$\Delta=\rho=\theta=\beta=0$.  The resulting dynamical equilibrium
initial condition is described by
\begin{eqnarray}
\zeta_{0}'&=&-\frac{\zeta_{0}}{\xi}-\beta_{0}\label{z0} \\
\beta_{0}'&=&-\frac{2\kappa_{0}}{p
\xi}\frac{\zeta_{0}}{\sqrt{\kappa_{1}^{2}-\zeta_{0}^{2}}}\label{b0} \\
\sigma_{0}&=&\sqrt{\kappa_{1}^{2}-\zeta^{2}_{0}}\label{wees0}\\
\Sigma_{0}&=&\kappa_{0}/\sigma_{0}\label{bigs0}
\end{eqnarray}
where $0$ subscripts denote equilibrium value and $\kappa_{0},
\kappa_{1}$ are constants.  The equilibrium equations describe the
self-consistent response of the plasma to a prescribed background
inhomogeneous magnetic field, $\beta_{0}$.  Conversely they describe
how the plasma must behave in order to generate the same magnetic
field.  Equation(\ref{wees0}) defines the kinetic energy
conservation of the equilibrium flow; equation (\ref{bigs0}) defines
the conservation of the total number density flux; and equations
(\ref{b0}),(\ref{z0}) describe the magnetic field generation via the
$\zeta_{0}$, $\beta_{0}$ coupling.

Linearise the full set of governing equations and look at large
values of $\xi$. In this regime the magnetic field is asymptotically
tending to a constant value, $\beta_{0}=constant$, corresponding to
no motion of the plasma in the azimuthal direction
$\zeta_{0}=\chi_{0}=0$, saying that radial flow dominates. This
requires $\sigma_{0}=constant$ to be consistent with equation
(\ref{wees0}) and implies that in this parameter set that there is a
net motion of the plasma in the radial direction with no net
current, $\delta_{0}=0$.  The governing equations then become,
\begin{eqnarray}
\dot{\Sigma}&=&-\left(\Sigma_{0}\sigma+\Sigma\sigma_{0}\right)' \label{bigs}\\
\dot{\Delta}&=&-\left(\sigma_{0}\Delta+\delta\Sigma_{0}\right)'\label{bigd} \\
\dot{\sigma}&=&-\sigma_{0}\sigma'+\beta_{0}\zeta \label{wees}\\
\dot{\delta}&=&-\sigma_{0}\delta'+\rho+\beta_{0}\chi \label{weed}\\
\dot{\chi}&=&-\sigma_{0}\chi'-\sigma_{0}\chi/\xi-\delta\beta_{0} \label{chi}\\
\dot{\zeta}&=&-\sigma_{0}\zeta'-\sigma_{0}\zeta/\xi-\sigma_{0}\beta-\beta_{0}\sigma+\theta\label{zeta}\\
\dot{\rho}&=&-\frac{2}{\xi}\left(\Delta\sigma_{0}+\Sigma_{0}\delta\right)\label{rho}\\
\dot{\theta}&=&-p\beta'-2\Sigma_{0}\zeta/\xi \label{theta}\\
\dot{\beta}&=&-\theta'-\theta/\xi\label{beta}
\end{eqnarray}

\section{Stability analysis}
Upon inspection of the linearised governing equations it is evident
that they can be split into two independent, self-consistent sets
namely equations (\ref{bigd}),(\ref{weed}),(\ref{chi}),(\ref{rho})
forming one set and equations
(\ref{bigs}),(\ref{wees}),(\ref{zeta}),(\ref{theta}) and
(\ref{beta}) forming the other.  This simplification was exploited
to obtain the following solutions.

\subsection{Electrostatic solution}
The electrostatic solution is characterised by $\beta=\beta_{0}$,
$\theta=0$ and $\zeta=0$.  Substituting these conditions into the
governing equations yields,
\begin{eqnarray}
\dot{\Delta}&=&-(\sigma_{0}\Delta+\delta\Sigma_{0})'\label{static1} \\
\dot{\delta}&=&-\sigma_{0}\delta'+\rho+\beta_{0}\chi \label{static2}\\
\dot{\chi}&=&-\sigma_{0}\chi'-\sigma_{0}\chi/\xi-\beta_{0}\delta \label{static3}\\
\dot{\rho}&=&-2(\Delta\sigma_{0}+\Sigma_{0}\delta)/\xi
\label{static4}
\end{eqnarray}
Combing equations (\ref{static2}) and (\ref{static3}) to eliminate
$\chi$ and (\ref{static4}) and (\ref{static1}) to eliminate $\Delta$
produces two differential equations both involving $\delta$ and
$\rho$.  Substituting one differential expression into the other to
eliminate $\rho$ yields the partial differential equation
\begin{eqnarray}
\ddot{\delta}+2\sigma_{0}\dot{\delta}'+\sigma_{0}^{2}\delta''+\sigma_{0}\dot{\delta}/\xi+\sigma_{0}^{2}\delta'/\xi
+\left(\beta_{0}^{2}+2\Sigma_{0}/\xi\right)\delta
\nonumber\\
-2\sigma_{0}C_{1}/\xi=0
\end{eqnarray}
where $C_{1}$ is a constant.  Setting $C_{1}=0$ and setting
$\delta=y(\xi)\exp{(-i\omega\tau)}$ yields
\begin{eqnarray}
\sigma_{0}^{2}y''+(\sigma_{0}^{2}/\xi-2i\omega\sigma_{0})y'+
(\beta_{0}^{2}+2\Sigma_{0}/\xi-\omega^{2} \nonumber
\\
-i\omega\sigma_{0}/\xi)y=0
\end{eqnarray}
This has complete solution
\begin{eqnarray}
\delta\left(\xi,\tau\right)=\xi^{-1/2}e^{i\omega(\xi/\sigma_{0}-\tau)}[C_{2}
M(-\frac{i\Sigma_{0}}{\sigma_{0}\beta_{0}},0,\frac{2i\beta_{0}\xi}{\sigma_{0}})
\nonumber
\\+C_{3}W(-\frac{i\Sigma_{0}}{\sigma_{0}\beta_{0}},0,\frac{2i\beta_{0}\xi}{\sigma_{0}})]\label{elecsol}
\end{eqnarray}
where $M,W$ are Whittaker functions of the first and second kind
respectively and $C_{2}, C_{3}$ are constants.  $M$ and $W$ are
related to the confluent hypergeometric functions $\mathcal{M}$ and
$\mathcal{U}$ as follows
\begin{eqnarray}
M(\kappa,\mu,z)&=&e^{-z/2}z^{1/2+\mu}\mathcal{M}(1/2+\mu-\kappa,1+2\mu,z)\\
W(\kappa,\mu,z)&=&e^{-z/2}z^{1/2+\mu}\mathcal{U}(1/2+\mu-\kappa,1+2\mu,z)
\end{eqnarray}
Taking into account the singularity at $\xi=0$ it is required that
the solution is bounded and well-behaved as $\xi\rightarrow0$.
Following Abramowitz and Stegun (\cite{abram}) for small arguments
$z\equiv 2i\beta_{0}\xi/\sigma_{0}$ the confluent hypergeometric
function of the second kind, $\mathcal{U}\propto\ln{z}\rightarrow
-\infty$ as $\xi\rightarrow 0$ implying $W$ is an unsuitable
solution. However as $\left|z\right|\rightarrow 0$ the function
$\mathcal{M}(a,b,0)=1$ provided $b$ is not a negative integer, in
this case $b=1$ implying $M$ is a bounded, well-behaved solution.
Hence we choose
\begin{equation}
\delta\left(\xi,\tau\right)=C_{2}\xi^{-1/2}e^{i\omega(\xi/\sigma_{0}-\tau)}
M(-\frac{i\Sigma_{0}}{\sigma_{0}\beta_{0}},0,\frac{2i\beta_{0}\xi}{\sigma_{0}})
\end{equation}
Rewriting $M$ as a series of modified Bessel functions of the first
kind, $I$,
\begin{eqnarray}
M(\kappa,\mu,z)=e^{-z/2}z^{1/2+\mu} \times \nonumber \\
\qquad\sum_{n=0}^{\infty}D_{n}(1/2+\mu-\kappa,1+2\mu)I_{n}(z)
\end{eqnarray}
 and using the asymptotic expansion of $I$, the solution takes the form
\begin{equation}
\delta(\xi,\tau)=(2\pi\xi)^{-1/2}C_{4}\exp{[i\omega(\xi/\sigma_{0}-\tau)+i\beta_{0}\xi/\sigma_{0}]}
\end{equation}
where $D$ is a set of real and imaginary coefficients and $C_{4}$ is
an arbitrary constant. In the uniform equilibrium situation the
electrostatic oscillation behaves as described in section $2.1$ at
$\omega^{2}=2\omega_{p}^{2}+\omega_{c}^{2}$ (Diver et al.
\cite{diver}); in dynamical equilibrium the electrostatic
oscillation is being convected at the plasma flow speed
$\sigma_{0}$.
\subsection{Convective solution}
If $\dot{\beta}=\delta=0$ and $\theta\sim1/\xi\neq\theta(\tau)$ this
yields
\begin{eqnarray}
\dot{\Sigma}&=&-(\Sigma_{0}\sigma+\Sigma\sigma_{0})' \label{con1}\\
\dot{\Delta}&=&-(\Sigma_{0}\Delta)'\label{con2}\\
\dot{\sigma}&=&0 \label{con2.5}\\
0&=&\rho+\beta_{0}\chi \label{con3}\\
\dot{\zeta}&=&0 \\
\dot{\chi}&=&-\sigma_{0}\chi'-\sigma_{0}\chi/\xi\label{con4} \\
\dot{\rho}&=&-2\Delta\sigma_{0}/\xi \label{con5}
\end{eqnarray}
Rearranging equation (\ref{con5}) for $\Delta$ and substituting into
(\ref{con2}) gives a differential expression for $\rho$.  The same
result can be achieved by rearranging (\ref{con3}) for $\chi$ and
substituting into (\ref{con4}).  Additionally integrating
(\ref{con1}) with respect to $\tau$ and using (\ref{con2.5}) gives a
differential equation describing the evolution of $\Sigma$.
Therefore,
\begin{eqnarray}
\dot{\Sigma}&=&-\sigma_{0}\Sigma'+C_{5}\\
\dot{\Delta}&=&-\sigma_{0}\Delta'\\
\dot{\chi}&=&-\sigma_{0}\left(\xi\chi\right)'/\xi \\
\dot{\rho}&=&-\sigma_{0}\left(\xi\rho\right)'/\xi
\end{eqnarray}
where $C_{5}$ is a constant.  These have the general solution,
\begin{eqnarray}
\Sigma(\xi,\tau)&=&f(\tau-\xi/\sigma_{0})-\xi C_{5}/\sigma_{0}\\
\Delta(\xi,\tau)&=&f(\tau-\xi/\sigma_{0})\\
G(\xi,\tau)&=&\frac{1}{\xi}f(\tau-\xi/\sigma_{0})
\end{eqnarray}
where $G=\rho,\chi$ and $f$ is an arbitrary functions.  Here
$\Sigma$, $G$ and $\Delta$ are being convected at the streaming
velocity of the plasma $\sigma_{0}$.  Assuming $G,\Delta,\Sigma \sim
\exp{(-i\omega\tau)}$ gives the particular solutions
$\propto\exp{[-i\omega(\tau-\xi/\sigma_{0})]}$ describing a
perturbation in the plasma variables propagating through the plasma
at the streaming velocity $\sigma_{0}$.
\subsection{General solution}
If we prescribe $\theta=f(\gamma)/\xi$ where $\gamma=k\xi-(\omega\pm
k\sigma_{0})\tau=k\xi-\Omega\tau$, looking at (\ref{beta}) this
implies $\beta=\eta f(\gamma)/\xi$, where $\eta$ is a constant.
Substituting this expression into (\ref{theta}) yields
\begin{equation}
\zeta=\frac{\left(\Omega-pk\eta\right)f_{\gamma}}{2\Sigma_{0}}+\frac{p\eta
f}{2 \Sigma_{0}\xi} \label{zetgen}
\end{equation}
Substituting (\ref{zetgen}) into (\ref{wees}) and further
prescribing $\sigma=\lambda f(\gamma)/\xi$, where $\eta$ is a
constant, requires
\begin{eqnarray}
(\sigma_{0}k-\Omega)\lambda/\xi&=&(\Omega-pk\eta)\beta_{0}/2\Sigma_{0}\\
-\sigma_{0}\lambda/\xi&=&p\beta_{0}\eta/2\Sigma_{0}
\end{eqnarray}
for $\sigma$, $\zeta$, $\theta$ and $\beta$ to be consistent.
Substituting the expressions for $\sigma$, $\zeta$, $\theta$ and
$\beta$ into equation (\ref{zeta}) defines the function $f$ by
\begin{eqnarray}
A\left(\sigma_{0}k-\Omega\right)f_{\gamma\gamma}
+\left[A\sigma_{0}+B\left(\sigma_{0}k-\Omega\right)\right]\frac{f_{\gamma}}{\xi}
\nonumber\\
-\left(1-\sigma_{0}\eta-\beta_{0}\lambda\right)\frac{f}{\xi}=0
\end{eqnarray}
where
\begin{eqnarray}
A&=&\left(\Omega-pk\eta\right)/2\Sigma_{0}\\
B&=&p\eta/2\Sigma_{0}
\end{eqnarray}
The solution of which is a Bessel function of non-integer order
which shows presence of Doppler effect as electromagnetic wave
propagates in the plasma that is streaming in the radial direction
with a velocity $\sigma_{0}$.
\section{Discussion and Further Developments}
The linear perturbation analysis presented here has shown some of
the dynamical responses of the magnetised electron-positron
streaming plasma. These offer guidance in respect of possible wave
modes and plasma stability.  In the linear regime the electrostatic,
convective and general solutions describe the simplified streaming
system perturbed from equilibrium.  The non-linear system of
equations does not have a closed-form analytical solution so the
problem has to be solved numerically, work on which is currently
still in progress.  The linear analysis presented here will help in
the development of the numerical simulations.  Future considerations
include extending the cold plasma treatment to a kinetic one.  In
this context the possibility exists of coupling Bernstein modes
(Laing \& Diver \cite{laing}) (electrostatic waves) to
electromagnetic modes.

These results imply a change in the study of pulsar radiation
mechanisms. In pulsar magnetospheres the electromagnetic field
distribution of the star is the superposition of the underlying
dipolar electromagnetic field of the star, plus the self-field of
the flowing plasma (da Costa \& Kahn \cite{costa82}, da Costa {\it
et al}. \cite{costa01}).  Collective processes in the pulsar rest
frame depend very strongly on the local plasma and field conditions.
Earlier work (Diver {\it et al}. \cite {diver}) explored quasilinear
coupling processes in a stationary inhomogeneous plasma. This
article presents the preliminary analysis of the stability of the
full non-linear mechanism in a streaming plasma prior to a full
numerical simulation.

\begin{acknowledgements}
We gratefully acknowledge the support by PPARC for a studentship for
C. R. Stark and the Barber Trust Fund.  We would also like to thank
the WE-Heraeus foundation for the opportunity to attend the
$363^{rd}$ Heraeus Seminar.
\end{acknowledgements}

%\end{document}
              \clearpage


\begin{thebibliography}{}
\bibitem[1965]{abram} Abramowitz, M., \& Stegun, I.A., 1965, Handbook of
Mathematical Functions (Dover Publications, Inc. New York)
\bibitem[1982]{costa82} da Costa, A.A., \& Kahn, F.D. 1982, MNRAS,
199, 211
\bibitem[1997]{costa97} da Costa, A.A., \& Kahn, F.D. 1997, MNRAS,
284, 1
\bibitem[2001]{costa01} da Costa, A.A., Diver, D.A., Stewart, G.A., 2001, A\&A 366,
129
\bibitem[2002]{diver} Diver, D.A., da Costa A.A., Laing, E.W., 2002, A\&A 387,
339
\bibitem[2005]{laing} Laing, E.W., \& Diver, D.A., 2005, Phys. Rev.
E 72, 036409
\end{thebibliography}
\end{document}